\documentclass[aps,prd,onecolumn,amsmath,11pt,superscriptaddress,floatfix,nofootinbib,preprintnumbers]{revtex4-2}

\usepackage{mathrsfs}
\usepackage{amsfonts}
\usepackage{amsmath,amssymb,bm}
\usepackage{array}
\usepackage{verbatim}
\usepackage{epsfig}
\usepackage{graphicx}
\usepackage{hyperref}
\hypersetup{colorlinks, linkcolor = [rgb]{0, 0, 0.5}, citecolor = [rgb]{0,0.0,0.5}, urlcolor = [rgb]{0,0.0,0.5}}
\usepackage[normalem]{ulem}
\usepackage{xcolor}
\usepackage{ulem}
\usepackage{multirow}

\usepackage{subcaption}
\usepackage{graphicx}
\graphicspath{{./figs/}}
\usepackage{dcolumn}
\usepackage{bm}
\usepackage{slashed}
\usepackage{siunitx}
\usepackage{ulem,xpatch}
\usepackage{hyperref}
\usepackage[mathlines]{lineno}
\usepackage{comment}
\usepackage{soul}
\usepackage{xcolor}
\usepackage{xfrac}

\allowdisplaybreaks[4]

\begin{document}

\title{On the Stability of Leading-Power Factorization under Photon Propagator Numerator Modifications}

\author{Cong Li}
\affiliation{School of Information Engineering, Zhejiang Ocean University, Zhoushan, Zhejiang, China}

\date{\today}


\begin{abstract}
We study collinear factorization in strong electromagnetic backgrounds within SCET for a class of modifications where the photon propagator keeps the vacuum pole structure and $i\varepsilon$ prescription, while the background enters only through a numerator tensor $\Delta_{\mu\nu}(k)$. We show that the set of Landau pinch surfaces and leading momentum regions is unchanged, so the leading-power (LP) factorized form is preserved. Moreover, the LP cusp kernel depends on the background solely through the longitudinal contraction $n^\mu\bar n^\nu\Delta_{\mu\nu}(k)$ in the soft region; if it vanishes (or is power suppressed), the LP soft kernel reduces to the vacuum. As an application, for an occupancy-number modification with the physical polarization-sum tensor $g_{T\mu\nu}(k;n)$, transversality implies $n^\mu\bar n^\nu\Delta_{\mu\nu}=0$, so genuine background sensitivity starts only beyond LP.
\end{abstract}

\maketitle

\section{Introduction}
\label{sec:introduction}
In ultraperipheral collisions (UPCs) \cite{BALTZ20081} and processes relevant to future electron--ion colliders (EICs) \cite{Accardi2016}, hard scattering often occurs in strong electromagnetic environments or non-vacuum photon states. In phenomenological analyses, one usually continues to use the collinear factorization formula established in the vacuum \cite{COLLINS1985199}, treating the background effect as a parametrizable external field or as a correction to the cross section from an equivalent photon distribution \cite{BUDNEV1975181}. A basic question is whether a background entering an internal gauge-boson propagator can modify the LP soft kernel. We show that it cannot, under the assumptions stated below.

A proof of collinear factorization relies on two independent inputs. First, the classification of pinch surfaces and leading momentum regions in loop integrals is determined by the analytic structure of propagator denominators and the $i\varepsilon$ prescription (Landau singularity analysis is particularly crucial in this respect \cite{LANDAU1959181,Coleman1965}). Second, at LP, soft radiation can be eikonalized and reorganized into Wilson-line matrix elements along ray directions, thereby systematically packaging long-distance soft physics into a soft function \cite{PhysRevD.8.4332}. We focus on a class of background modifications whose common feature is that the pole structure and prescription of the propagator remain in the vacuum form, where the denominator is still $k^2+i\varepsilon$, and the background dependence enters only through the numerator tensor structure.

Under this setup, we prove that the form of factorization at LP remains unchanged: the organization of hard matching and collinear sectors is the same as in the vacuum, and any possible LP background dependence can only enter through the definition of soft matrix elements, i.e.\ through the background soft function $S^{\mathrm{bg}}=\langle W_n^{(s)} W_{\bar n}^{(s)\dagger}\rangle_{\mathrm{bg}}$.

We provide a criterion to determine when the LP soft kernel reduces to the vacuum result. The cusp-related LP soft kernel depends on the background only through the longitudinal contraction along the two eikonal directions \cite{KORCHEMSKY1987342}. We apply the criterion to a concrete model: we regard the background as a mixed state described by the occupation number of real photons $f(k)$, and use the physical polarization-sum tensor $g_{T\mu\nu}(k;n)$ to represent its numerator structure (with the axial-gauge reference vector chosen as $r=n$). In this model, transversality implies that the LP soft kernel is unchanged; genuine background sensitivity can only start from subleading power, for example through dynamical corrections with $p_\perp\sim Q\lambda$ or transverse couplings induced by subleading SCET Lagrangians~\cite{PhysRevD.63.114020}.

The structure of this paper is as follows. Section~\ref{sec:propa_bg} derives, starting from the density matrix of the background state, the propagator modification induced by the occupation number of real photons and its tensor structure. Section~\ref{sec:fact_bg} proves within SCET the stability of the LP factorized form when the denominator pole structure is kept unchanged, and provides the criterion for the LP soft kernel to reduce to the vacuum. Section~\ref{sec:soft_factor_bg} discusses the exponentiation of the background soft function and applies the criterion to the concrete model of Sec.~\ref{sec:propa_bg}. Finally, the conclusions summarize the results and indicate possible manifestations of background effects at subleading power.

\section{Photon propagator modified by the background field}
\label{sec:propa_bg}
Before going into specific models in detail, it is first necessary to clarify the physical domain of applicability for the background propagator modification adopted in this work. In general, the photon propagator in a medium at finite temperature or finite density receives two types of modifications \cite{LANDSMAN1987141}: (i) the real part of the self-energy modifies the dispersion relation, such as generating a thermal mass (TM) or Debye screening (DS), which changes the pole position in the denominator (e.g., $k^2 \to k^2 - m_D^2$) \cite{PhysRevD.26.1394}; (ii) the imaginary part of the self-energy or changes in statistical weights induce modifications to the numerator tensor structure \cite{PhysRevD.9.3320}. In this work, we assume that the characteristic soft scale probed by the hard scattering is much larger than the background-induced mass scale ($k_{\mathrm{soft}}^2 \gg m_D^2$), so that Debye-mass effects can be regarded as higher-order and power suppressed. Under this approximation, we keep the denominator pole structure in the vacuum form $k^2+i\varepsilon$, and focus on how medium-induced numerator modifications from background particle occupation numbers affect the stability of the factorization structure.

As is well known, in the vacuum, the photon propagator is $\langle 0 |T A_\mu (x) A_\nu (y) | 0 \rangle$. However, when the photon propagator is in an external field, the background field is $B$, which is a mixed state composed of real photons; for example, in heavy-ion collisions, $B$ corresponds to the photon field around the heavy ions. The photon propagator consists of two parts:
\begin{align}
\langle 0 | TA_\mu (x) A_\nu (y) | 0\rangle+\langle B | TA_\mu (x) A_\nu (y) | B \rangle.
\end{align}
The first term is the familiar vacuum propagator, and the second term arises from the background real-photon field. Furthermore, we have
\begin{align}
\langle B |T A_\mu (x) A_\nu (y) | B \rangle
&= \text{Tr}[\rho TA_\mu (x) A_\nu (y)] \notag\\
&= \sum_{\lambda} \int \frac{d^3 k}{(2\pi)^3 2k_0} f(k) \text{Tr}[ |k\rangle \langle k|T A_\mu (x) A_\nu (y)] \notag\\
&= \sum_{\lambda} \int \frac{d^3 k}{(2\pi)^3 2k_0} f(k) \langle k |T A_\mu (x) A_\nu (y) | k \rangle \notag\\
&= \int \frac{d^3 k}{(2\pi)^3 2k_0} f(k) e^{-ik\cdot(x-y)} \theta(x^0-y^0)\sum_{\lambda} \epsilon_\mu^{\lambda*} \epsilon_\nu^\lambda+(x\leftrightarrow y) \notag\\
&= \int \frac{d^3 k}{(2\pi)^3 2k_0} f(k) e^{-ik\cdot(x-y)}\theta(x^0-y^0) (-g_{T\mu\nu})+(x\leftrightarrow y) \notag\\
&= \int \frac{d^4 k}{(2\pi)^4} e^{-ik\cdot(x-y)} \frac{-ig_{T\mu\nu}}{k^2+i\varepsilon} f(k)
\label{pro}
\end{align}
where $A_\nu (y) | k \rangle=\epsilon^\lambda_\nu(k) e^{-ik\cdot y} | 0 \rangle$ and
$$
\rho = | B \rangle \langle B | = \sum_{\lambda} \int \frac{d^3 k}{(2\pi)^3 2k_0} f(k) | k \rangle \langle k |
$$
is the spin density matrix of the background field \cite{NIEMI1984105,Kobes1985}. In the last line of Eq.~(\ref{pro}), we use the standard residue theorem to combine the two time-ordered terms into $\frac{1}{k^2+i \varepsilon}$, in which $f(k)$ is an even function. Here $f(k)$ is the average occupation number of photons, representing the number of real photons with momentum magnitude $|k|=E_k$. One can see that the photon propagator in a real-photon field differs from that in the vacuum: it is the vacuum propagator convoluted with an occupation-number function $f(k)$, indicating that photon exchange is influenced by the absolute number of real photons. Therefore, the propagator in the background field can be written as
\begin{align}
D^{bg}_{\mu\nu}(x,y)&=\langle 0 | TA_\mu (x) A_\nu (y) | 0 \rangle+\langle B | TA_\mu (x) A_\nu (y) | B \rangle \notag\\
&= \int \frac{d^4 k}{(2\pi)^4} e^{-ik\cdot(x-y)} \frac{-ig_{\mu\nu}}{k^2+i\varepsilon}+\int \frac{d^4 k}{(2\pi)^4} e^{-ik\cdot(x-y)} \frac{-ig_{T\mu\nu}}{k^2+i\varepsilon} f(k)
\label{proo}
\end{align}
Such a propagator leads to the following physical picture: when the scattering between two electrons takes place in a background field with a photon distribution $f(k)$, they can scatter not only through virtual photons in the vacuum, but also through real photons in the background field. Finally, we emphasize that the polarization-vector sum for real photons is transverse to the photon direction of motion, and we have
\begin{equation}
-g_{T\mu\nu}=\sum\epsilon^{\lambda\,*}_\mu(k)\,\epsilon^{\lambda}_\nu(k)=-\,g_{\perp\mu\nu}+\frac{k_{\perp\mu} n_\nu+k_{\perp\nu} n_\mu}{n\cdot k}-\frac{k_\perp^2}{(n\cdot k)^2}\,n_\mu n_\nu,
\label{eq:Pmunu_axial_n_expanded}
\end{equation}
which differs from $-g_{\perp\mu\nu}=-g_{\mu\nu}+\bar{n}_\mu n_\nu/2+\bar{n}_\nu n_\mu/2$ used below. The light-cone vectors $n,\bar n$ satisfy $n^2=\bar n^2=0$, $n\!\cdot\!\bar n=2$. Regarding the spurious pole $n\!\cdot\!k=0$ in $g_T$ under the axial-gauge representation, Eq.~\eqref{eq:Pmunu_axial_n_expanded} writes the physical polarization-sum tensor $g_{T\mu\nu}(k;n)$ in an explicit axial-gauge form (with reference vector $n$), hence the factors $1/(n\!\cdot\!k)$ and $1/(n\!\cdot\!k)^2$. These factors are not new dynamical propagator denominators, but spurious poles introduced by the gauge choice. To preserve causal analyticity and the standard contour deformation in energy integrals, we adopt by default the Mandelstam--Leibbrandt (ML) prescription to define them \cite{MANDELSTAM1983149,RevModPhys.59.1067}, for which we may take
\begin{equation}
\frac{1}{n\!\cdot\!k}\Big|_{\rm ML}\equiv\lim_{\eta\to0^+}\frac{\bar n\!\cdot\!k}{(n\!\cdot\!k)(\bar n\!\cdot\!k)+i\eta},
\label{eq:ML_prescription}
\end{equation}
thereby avoiding mistaking $n\!\cdot\!k=0$ as a new physical pole that would generate an additional pinching mechanism. More importantly, the LP soft kernel discussed later enters only through the longitudinal eikonal projection, i.e., it depends only on $n^\mu\bar n^\nu D^{\rm bg}_{\mu\nu}(k)$. The tensor structure $g_{T\mu\nu}(k;n)$ satisfies transversality; although one representation of $g_{T\mu\nu}$ contains spurious poles, they do not contribute in the LP soft kernel studied here, and thus do not alter the classification of Landau singularities/leading regions determined by the physical denominator $k^2+i\varepsilon$.

\section{Stability of leading regions and LP factorization form under background-induced numerator modifications}
\label{sec:fact_bg}
We consider a class of background modifications in which an internally exchanged photon propagator keeps the vacuum denominator and $i\varepsilon$ prescription, while the background enters only through a numerator tensor $\Delta_{\mu\nu}(k)$. The stability of the LP factorization form hinges on the persistence of the vacuum-like pinch surfaces. Since the modification in Eq.(\ref{proo}) preserves the denominator $k^2+i\epsilon$, the set of leading momentum regions is identical to the vacuum case. We will show that, provided the background introduces no additional physical analytic singularities or new momentum scales beyond the vacuum pole structure, the candidate set of leading regions is unchanged and the LP factorized form is preserved. Under these conditions, any possible LP background dependence can enter only through the soft matrix element. Whether the LP soft kernel itself reduces to the vacuum result in a given model will be addressed in Sec.~\ref{sec:soft_factor_bg}.

\subsection{Setup and power counting}
\label{subsec:setup_assump}
We parameterize the background-modified momentum-space propagator as
\begin{equation}
D_{\mu\nu}^{\mathrm{bg}}(k)=\frac{-i}{k^2+i\varepsilon}\Big[g_{\mu\nu}+\Delta_{\mu\nu}(k)\Big].
\label{eq:Dbg_general_prd3_re}
\end{equation}
Throughout this section we impose:
(i) the denominator remains $k^2+i\varepsilon$ (no new physical poles/branch cuts or additional scales);
(ii) $\Delta_{\mu\nu}(k)$ does not introduce additional \emph{physical} analytic singularities in the momentum regions relevant for factorization.
Apparent singularities in specific representations (e.g.\ axial-gauge projectors) are treated as spurious and are assumed to cancel after physical contractions (Ward identities/current conservation) once a consistent prescription is adopted.

We use standard SCET power counting with hard scale $Q$ and $\lambda\ll 1$,
\begin{equation}
n\text{-collinear}:~p^\mu\sim Q(1,\lambda^2,\lambda),\qquad
\bar n\text{-collinear}:~p^\mu\sim Q(\lambda^2,1,\lambda),\qquad
\text{soft}:~k^\mu\sim Q(\lambda,\lambda,\lambda),
\label{eq:scet_scaling_prd3_re}
\end{equation}
with $p^\mu=(n\!\cdot\!p,\bar n\!\cdot\!p,p_\perp)$ and $n\!\cdot\!\bar n=2$ \cite{Bauer_2002,Beneke:2002ph}. Since the soft and collinear modes have the same virtuality scaling (SCET$_{\rm II}$), soft functions are understood as rapidity-regulated objects; we always compare vacuum and background within the same rapidity scheme \cite{PhysRevD.76.074002,PhysRevLett.108.151601,ECHEVARRIA2013795,BECHER201241}.

\subsection{Stability of pinch surfaces and leading momentum regions}
\label{subsec:regions_hard}
Leading regions are controlled by the analytic structure of propagator denominators and their $i\varepsilon$ prescriptions \cite{LANDAU1959181,Coleman1965,PhysRevD.18.3252}. With the propagator written as in Eq.~\eqref{eq:Dbg_general_prd3_re}, the physical denominator remains $k^2+i\varepsilon$, hence the pole locations in energy integrals are identical to those in the vacuum. Therefore, the set of candidate pinch surfaces and the classification of leading momentum regions (hard/collinear/soft and overlaps) are not enlarged by the background.

This does not preclude background effects at LP: $\Delta_{\mu\nu}(k)$ can reweight tensor contractions within existing regions and can demote (or eliminate) contributions that are leading in the vacuum. We assume standard current conservation/Ward identities for the external states so that LP soft interactions eikonalize and can be organized into Wilson lines \cite{PhysRev.140.B516,Bauer_2002,Beneke:2002ph}.

Hard matching is localized to the hard region as usual \cite{BENEKE1998321}. Background sensitivity in the hard coefficient depends on the behavior of $\Delta_{\mu\nu}(k)$ for $k\sim Q$: if $\Delta_{\mu\nu}(k)$ has support mainly at soft/semi-soft scales or admits a smooth expansion at hard momenta, its effect in hard matching is power suppressed and can be absorbed into subleading operators,
\begin{equation}
C^{\mathrm{bg}}(Q,\mu)=C^{\mathrm{vac}}(Q,\mu)+\mathcal{O}(\lambda).
\label{eq:hard_coeff_same_prd3_re}
\end{equation}
In UPC/EPA applications, the photon spectrum $f(k)$ is governed by the nuclear form factor and is strongly suppressed for $|k|\gg 1/R_A\ll Q$, so insertions of $\Delta_{\mu\nu}(k)$ into the hard region are power suppressed.

\subsection{LP factorization form and where background dependence can enter}
\label{subsec:lp_fact_form}
At LP, soft emissions off energetic charged particles eikonalize and are encoded in Wilson lines \cite{PhysRevD.8.4332}. Under the assumptions above, the LP factorized \emph{form} therefore remains the vacuum one,
\begin{equation}
\mathcal{M}\sim H(Q,\mu)\;J_n(\mu)\;J_{\bar n}(\mu)\;S(\mu),
\label{eq:factorization_form_prd3_re}
\end{equation}
with the only possible LP entry of background information being the definition of the soft matrix element as a Wilson-line expectation value in the background state,
\begin{equation}
S^{\mathrm{bg}}\equiv\Big\langle\, W_n^{(s)}(0)\,W_{\bar n}^{(s)\dagger}(0)\,\Big\rangle_{\mathrm{bg}}.
\label{eq:Sbg_def_prd3_re}
\end{equation}
Whether $S^{\mathrm{bg}}$ at LP reduces to the vacuum soft function or carries genuine background dependence is a dynamical question addressed in Sec.~\ref{sec:soft_factor_bg}, where we exponentiate $S^{\mathrm{bg}}$ and derive a practical LP projection criterion for the soft kernel, which we then apply to the occupancy-number model of Sec.~\ref{sec:propa_bg}.

\section{Background soft function, exponentiation, and the LP projection criterion}
\label{sec:soft_factor_bg}
Section~\ref{sec:fact_bg} established two facts under the class of background modifications in Eq.~(\ref{eq:Dbg_general_prd3_re}): (i) the set of Landau pinch surfaces and leading momentum regions is not enlarged because the physical denominator $k^2+i\varepsilon$ is unchanged; (ii) the LP factorized \emph{form} is preserved, with the only possible LP entry of background information being the soft matrix element $S^{\mathrm{bg}}$ defined by Wilson lines. The remaining question is then operational: how does one make the background soft function explicit in a way that reveals which tensor projections of $\Delta_{\mu\nu}(k)$ can affect the LP soft kernel, and when the LP soft kernel reduces to the vacuum result? In this section we answer this constructively by exponentiating the background soft function. This representation shows that the cusp-related LP soft kernel depends on the propagator tensor structure only through a single longitudinal contraction, yielding a simple LP projection criterion. More generally, the relation between Wilson-line correlators and the (cusp/soft) anomalous-dimension structure of infrared singularities is standard; see Ref.~\cite{PhysRevLett.102.162001}.
 We then apply it to the occupancy-number model derived in Sec.~\ref{sec:propa_bg}.

\subsection{Background soft function, cumulant expansion, and exponentiation}
\label{subsec:soft_exp_bg}
At LP in SCET, the interaction between energetic charged particles and soft photons eikonalizes and is described by Wilson lines along classical ray directions. For QED with two leading directions $n$ and $\bar n$, we introduce the soft Wilson lines
\begin{equation}
W_n^{(s)}(0)=\mathcal{P}\exp\!\left[\,ie\int_{-\infty}^{0}\!ds\,n\!\cdot\!A_s(sn)\right],\qquad
W_{\bar n}^{(s)}(0)=\mathcal{P}\exp\!\left[\,ie\int_{-\infty}^{0}\!ds\,\bar n\!\cdot\!A_s(s\bar n)\right],
\end{equation}
which resum arbitrarily many soft emissions along the two lightlike trajectories. The corresponding background soft function is defined as
\begin{equation}
S^{\mathrm{bg}}\equiv\left\langle\,W_n^{(s)}(0)\,W_{\bar n}^{(s)\dagger}(0)\,\right\rangle_{\mathrm{bg}}.
\label{eq:Sbg_def_sec4_final}
\end{equation}
To isolate propagator-induced effects and avoid additional classical phases, we assume $\langle A_\mu(x)\rangle_{\mathrm{bg}}=0$, since the measured cross section is effectively averaged over unobserved phases/orientations (e.g. impact parameter and recoil), which washes out event-by-event coherence and thus sets $\langle A_\mu\rangle_{\rm bg}=0$ while keeping $\langle T A_\mu A_\nu\rangle_{\rm bg}\neq 0$. All dynamical information entering Eq.~(\ref{eq:Sbg_def_sec4_final}) is then encoded in background-state correlation functions of the soft photon field, in particular the time-ordered two-point function
\begin{equation}
\langle T\,A_\mu(x)A_\nu(y)\rangle_{\mathrm{bg}}\equiv D^{\mathrm{bg}}_{\mu\nu}(x-y),
\end{equation}
whose momentum-space form is precisely Eq.~(\ref{eq:Dbg_general_prd3_re}).

The Wilson-line correlator in Eq.~(\ref{eq:Sbg_def_sec4_final}) can be written as a generating functional in the presence of classical sources that represent the eikonal currents carried by the energetic particles. Introducing the eikonal current
\begin{equation}
J^\mu(x)=e\,n^\mu\!\int_{-\infty}^{0}\!ds\,\delta^{(4)}(x-sn)-e\,\bar n^\mu\!\int_{-\infty}^{0}\!ds\,\delta^{(4)}(x-s\bar n),
\end{equation}
the logarithm of the soft function admits a cumulant (connected-diagram) expansion. This linked-cluster structure is general and holds for an arbitrary background state \cite{GATHERAL198390,FRENKEL1984231}. If one further adopts the Gaussian (two-point-only) approximation—namely, retaining only the connected two-point function of the background field—then all connected contributions to $\ln S^{\mathrm{bg}}$ are fixed by $D^{\mathrm{bg}}_{\mu\nu}$. In this case the cumulant expansion truncates at second order and yields \cite{YENNIE1961379}
\begin{equation}
\ln S^{\mathrm{bg}}\simeq-\frac12\int d^4x\,d^4y\;J^\mu(x)\,D^{\mathrm{bg}}_{\mu\nu}(x-y)\,J^\nu(y).
\label{eq:lnS_JDJ_sec4_final}
\end{equation}
Eq.~(\ref{eq:lnS_JDJ_sec4_final}) makes explicit that exponentiation of the soft factor is preserved in the presence of the background: the geometry resides entirely in the eikonal current $J^\mu$, while the background enters solely through the soft two-point function. In particular, under our setup the background does not modify the causal analyticity associated with the physical denominator $k^2+i\varepsilon$, and the resummation mechanism inherent in Wilson lines remains intact.

\subsection{Momentum-space kernel and the LP projection criterion}
\label{subsec:kernel_sec4_final}
To expose the LP structure of the exponent kernel, it is convenient to work in momentum space. Fourier transforming the eikonal current yields
\begin{equation}
J^\mu(k)=e\left(\frac{n^\mu}{n\!\cdot\!k+i\varepsilon}-\frac{\bar n^\mu}{\bar n\!\cdot\!k-i\varepsilon}\right),
\end{equation}
where the $i\varepsilon$ prescriptions originate from the convergence factors of the semi-infinite Wilson lines and are consistent with time ordering. Substituting into Eq.~(\ref{eq:lnS_JDJ_sec4_final}), the logarithm of the soft function becomes a sum of momentum-space kernels. Among these, the term corresponding to soft exchange between the two Wilson lines (the cusp contribution) reads
\begin{equation}
\ln S^{\mathrm{bg}}_{\mathrm{cross}}\simeq-\,e^2\int\frac{d^4k}{(2\pi)^4}\;
\frac{n^\mu\bar n^\nu\,D^{\mathrm{bg}}_{\mu\nu}(k)}{(n\!\cdot\!k+i\varepsilon)(\bar n\!\cdot\!k-i\varepsilon)}.
\label{eq:lnS_cross_sec4_final}
\end{equation}
In contrast, the remaining pieces in $\ln S^{\mathrm{bg}}$ correspond to self-energy–type contributions involving a single Wilson line; they contribute only to individual-line renormalization and overall phases and do not encode inter-jet soft correlations. Thus, the cusp-related LP soft kernel is fully controlled by the numerator contraction $n^\mu\bar n^\nu D^{\mathrm{bg}}_{\mu\nu}(k)$ in Eq.~(\ref{eq:lnS_cross_sec4_final}).

Using the decomposition in Eq.~(\ref{eq:Dbg_general_prd3_re}), we have
\begin{equation}
n^\mu\bar n^\nu\,D^{\mathrm{bg}}_{\mu\nu}(k)=\frac{-i}{k^2+i\varepsilon}\Big[n\!\cdot\!\bar n+n^\mu\bar n^\nu\,\Delta_{\mu\nu}(k)\Big].
\label{eq:nnbDbg_general_prd3_re}
\end{equation}
Eq.~(\ref{eq:nnbDbg_general_prd3_re}) makes the LP projection structure manifest: at LP, the cusp kernel is sensitive to the background-induced tensor modification \emph{only} through the single longitudinal contraction $n^\mu\bar n^\nu\Delta_{\mu\nu}(k)$ evaluated in the soft region. This yields an operational LP criterion:
\begin{equation}
n^\mu\bar n^\nu\,\Delta_{\mu\nu}(k)=0\qquad(\text{or power suppressed for }k\sim Q\lambda)\ \Longrightarrow\ S^{\mathrm{bg}}_{\mathrm{LP}}=S^{\mathrm{vac}}_{\mathrm{LP}}.
\label{eq:criterion_soft_vac_prd3_re}
\end{equation}
Equivalently, if Eq.~(\ref{eq:criterion_soft_vac_prd3_re}) holds, the LP cross kernel in Eq.~(\ref{eq:lnS_cross_sec4_final}) reduces identically to the vacuum one, and therefore
\begin{equation}
S^{\mathrm{bg}}_{\mathrm{LP}}=S^{\mathrm{vac}}_{\mathrm{LP}}.
\label{eq:Sbg_to_Svac_prd3_re}
\end{equation}
Conversely, if Eq.~(\ref{eq:criterion_soft_vac_prd3_re}) does not hold, the factorized \emph{form} established in Sec.~\ref{sec:fact_bg} still holds, but the LP soft kernel generally carries background dependence through $n^\mu\bar n^\nu\Delta_{\mu\nu}(k)$ in the soft region. Importantly, this conclusion is independent of any particular gauge representation of $\Delta_{\mu\nu}$: it follows purely from the eikonal numerators and the geometry of the two Wilson-line directions.

\subsection{Application: occupancy-number model and explicit realization of the LP criterion}
\label{subsec:app_occ}
We now apply the criterion in Eq.~(\ref{eq:criterion_soft_vac_prd3_re}) to the occupancy-number modification derived in Sec.~\ref{sec:propa_bg}. In that model, the background modification takes the form
\begin{equation}
\Delta_{\mu\nu}(k)=g_{T\mu\nu}(k;n)\,f(k),
\label{eq:Delta_gTfk_prd3_re}
\end{equation}
where $f(k)$ is the average occupation number of real photons and $g_{T\mu\nu}(k;n)$ is the physical polarization-sum tensor defined with a reference vector chosen as the light-cone direction $n$. Its key property is transversality with respect to the Wilson-line direction,
\begin{equation}
n^\mu g_{T\mu\nu}(k;n)=0,\qquad k^\mu g_{T\mu\nu}(k;n)=0,
\label{eq:gT_transverse_prd3_re}
\end{equation}
which implies
\begin{equation}
n^\mu\bar n^\nu\,\Delta_{\mu\nu}(k)=f(k)\,n^\mu\bar n^\nu\,g_{T\mu\nu}(k;n)
=f(k)\,\bar n^\nu\big(n^\mu g_{T\mu\nu}(k;n)\big)=0.
\label{eq:corollary_gT_prd3_re}
\end{equation}
Therefore, the occupancy-number model automatically satisfies the LP criterion in Eq.~(\ref{eq:criterion_soft_vac_prd3_re}), and we conclude that$S^{\mathrm{bg}}_{\mathrm{LP}}=S^{\mathrm{vac}}_{\mathrm{LP}}$.

This result does not require approximating $g_{T\mu\nu}$ by $g_{\perp\mu\nu}$, nor does it rely on any accidental projection: it follows from a basic geometric fact that the LP eikonal current contains only the two longitudinal directions $n^\mu$ and $\bar n^\mu$, and any numerator modification transverse to the Wilson-line direction is eliminated in the LP longitudinal contraction. If a particular axial-gauge representation of $g_{T\mu\nu}(k;n)$ contains spurious poles (such as $1/(n\!\cdot\!k)$), they are gauge artifacts; in the LP cusp kernel they are projected out by Eq.~(\ref{eq:corollary_gT_prd3_re}), and therefore do not introduce new physical pinching beyond the denominator $k^2+i\varepsilon$.

The above conclusion is at LP. Because a realistic collinear momentum contains $p_\perp^\mu\sim Q\lambda$, transverse components start to participate in contractions at next-to-leading power (NLP); equivalently, in SCET there appear subleading couplings to $A_{s\perp}$ and insertions of subleading Lagrangians \cite{PhysRevD.63.114020,PhysRev.110.974,PhysRevLett.20.86}. Thus, even when $n^\mu\Delta_{\mu\nu}(k)=0$, the background tensor structure can enter observables at higher orders through transverse contractions and NLP operator insertions. In this work we focus on the LP projection criterion and its explicit realization; a systematic NLP analysis is left for future study.

\section{Conclusion}
\label{sec:conclusion}
We studied collinear factorization in strong electromagnetic backgrounds for a class of propagator deformations in which the photon denominator and $i\varepsilon$ prescription remain the vacuum ones, while background information enters only through a numerator tensor $\Delta_{\mu\nu}(k)$. Under the assumption that $\Delta_{\mu\nu}$ introduces no additional physical analytic singularities or new momentum scales, the Landau pinch surfaces and the set of candidate leading momentum regions are identical to the vacuum, and therefore the LP factorized form is preserved.

At LP, any background dependence can enter only through the soft matrix element defined by Wilson lines in the background state. By exponentiating this background soft function (under the Gaussian/two-point approximation), we showed that the cusp-related LP soft kernel depends on the propagator modification only through a single longitudinal contraction in the soft region, $n^\mu\bar n^\nu\Delta_{\mu\nu}(k)$. This yields a simple operational criterion: if $n^\mu\bar n^\nu\Delta_{\mu\nu}(k)$ vanishes (or is power suppressed for $k\sim Q\lambda$), then the LP soft kernel reduces to the vacuum and $S^{\mathrm{bg}}_{\mathrm{LP}}=S^{\mathrm{vac}}_{\mathrm{LP}}$.

As an application, for the occupancy-number modification $\Delta_{\mu\nu}(k)=g_{T\mu\nu}(k;n)\,f(k)$, transversality implies $n^\mu\bar n^\nu\Delta_{\mu\nu}=0$, so genuine background sensitivity starts beyond LP. The earliest contributions are expected at NLP, where transverse momenta and subleading SCET interactions allow transverse contractions and operator insertions that are absent at LP.

\section*{Acknowledgments}
\bibliography{ref}

@article{BALTZ20081,
title = {The physics of ultraperipheral collisions at the LHC},
journal = {Physics Reports},
volume = {458},
number = {1},
pages = {1-171},
year = {2008},
issn = {0370-1573},
doi = {https://doi.org/10.1016/j.physrep.2007.12.001},
url = {https://www.sciencedirect.com/science/article/pii/S0370157307004462},
author = {A.J. Baltz and G. Baur and D. d’Enterria and L. Frankfurt and F. Gelis and V. Guzey and K. Hencken and Yu. Kharlov and M. Klasen and S.R. Klein and V. Nikulin and J. Nystrand and I.A. Pshenichnov and S. Sadovsky and E. Scapparone and J. Seger and M. Strikman and M. Tverskoy and R. Vogt and S.N. White and U.A. Wiedemann and P. Yepes and M. Zhalov},
}

@article{BUDNEV1975181,
title = {The two-photon particle production mechanism. Physical problems. Applications. Equivalent photon approximation},
journal = {Physics Reports},
volume = {15},
number = {4},
pages = {181-282},
year = {1975},
issn = {0370-1573},
doi = {https://doi.org/10.1016/0370-1573(75)90009-5},
url = {https://www.sciencedirect.com/science/article/pii/0370157375900095},
author = {V.M. Budnev and I.F. Ginzburg and G.V. Meledin and V.G. Serbo},
}

@article{PhysRevD.63.114020,
  title = {An effective field theory for collinear and soft gluons: Heavy to light decays},
  author = {Bauer, Christian W. and Fleming, Sean and Pirjol, Dan and Stewart, Iain W.},
  journal = {Phys. Rev. D},
  volume = {63},
  issue = {11},
  pages = {114020},
  numpages = {17},
  year = {2001},
  month = {May},
  publisher = {American Physical Society},
  doi = {10.1103/PhysRevD.63.114020},
  url = {https://link.aps.org/doi/10.1103/PhysRevD.63.114020}
}

@article{Accardi2016,
  author  = {Accardi, A. and others},
  title   = {Electron-Ion Collider: The next QCD frontier},
  journal = {The European Physical Journal A},
  year    = {2016},
  volume  = {52},
  number  = {9},
  pages   = {268},
  doi     = {10.1140/epja/i2016-16268-9},
  url     = {https://doi.org/10.1140/epja/i2016-16268-9}
}

@article{LANDAU1959181,
title = {On analytic properties of vertex parts in quantum field theory},
journal = {Nuclear Physics},
volume = {13},
number = {1},
pages = {181-192},
year = {1959},
issn = {0029-5582},
doi = {https://doi.org/10.1016/0029-5582(59)90154-3},
url = {https://www.sciencedirect.com/science/article/pii/0029558259901543},
author = {L.D. Landau}
}

@article{Coleman1965,
  author  = {Coleman, S. and Norton, R. E.},
  title   = {Singularities in the physical region},
  journal = {Il Nuovo Cimento (1955-1965)},
  year    = {1965},
  volume  = {38},
  number  = {1},
  pages   = {438--442},
  doi     = {10.1007/BF02750472},
  url     = {https://doi.org/10.1007/BF02750472},
  issn    = {1827-6121}
}

@article{PhysRevD.18.3252,
  title = {Jet and lepton-pair production in high-energy lepton-hadron and hadron-hadron scattering},
  author = {Libby, Stephen B. and Sterman, George},
  journal = {Phys. Rev. D},
  volume = {18},
  issue = {9},
  pages = {3252--3268},
  numpages = {0},
  year = {1978},
  month = {Nov},
  publisher = {American Physical Society},
  doi = {10.1103/PhysRevD.18.3252},
  url = {https://link.aps.org/doi/10.1103/PhysRevD.18.3252}
}

@article{COLLINS1985199,
title = {Transverse momentum distribution in Drell-Yan pair and W and Z boson production},
journal = {Nuclear Physics B},
volume = {250},
number = {1},
pages = {199-224},
year = {1985},
issn = {0550-3213},
doi = {https://doi.org/10.1016/0550-3213(85)90479-1},
url = {https://www.sciencedirect.com/science/article/pii/0550321385904791},
author = {J.C. Collins and Davison E. Soper and George Sterman},
}

@article{MANDELSTAM1983149,
title = {Light-cone superspace and the ultraviolet finiteness of the N=4 model},
journal = {Nuclear Physics B},
volume = {213},
number = {1},
pages = {149-168},
year = {1983},
issn = {0550-3213},
doi = {https://doi.org/10.1016/0550-3213(83)90179-7},
url = {https://www.sciencedirect.com/science/article/pii/0550321383901797},
author = {Stanley Mandelstam},

}

@article{RevModPhys.59.1067,
  title = {Introduction to noncovariant gauges},
  author = {Leibbrandt, George},
  journal = {Rev. Mod. Phys.},
  volume = {59},
  issue = {4},
  pages = {1067--1119},
  numpages = {0},
  year = {1987},
  month = {Oct},
  publisher = {American Physical Society},
  doi = {10.1103/RevModPhys.59.1067},
  url = {https://link.aps.org/doi/10.1103/RevModPhys.59.1067}
}

@article{YENNIE1961379,
title = {The infrared divergence phenomena and high-energy processes},
journal = {Annals of Physics},
volume = {13},
number = {3},
pages = {379-452},
year = {1961},
issn = {0003-4916},
doi = {https://doi.org/10.1016/0003-4916(61)90151-8},
url = {https://www.sciencedirect.com/science/article/pii/0003491661901518},
author = {D.R Yennie and S.C Frautschi and H Suura},
}

@article{KORCHEMSKY1987342,
title = {Renormalization of the Wilson loops beyond the leading order},
journal = {Nuclear Physics B},
volume = {283},
pages = {342-364},
year = {1987},
issn = {0550-3213},
doi = {https://doi.org/10.1016/0550-3213(87)90277-X},
url = {https://www.sciencedirect.com/science/article/pii/055032138790277X},
author = {G.P. Korchemsky and A.V. Radyushkin},
}

@article{LANDSMAN1987141,
title = {Real- and imaginary-time field theory at finite temperature and density},
journal = {Physics Reports},
volume = {145},
number = {3},
pages = {141-249},
year = {1987},
issn = {0370-1573},
doi = {https://doi.org/10.1016/0370-1573(87)90121-9},
url = {https://www.sciencedirect.com/science/article/pii/0370157387901219},
author = {N.P. Landsman and Ch.G. {van Weert}},

}

@article{PhysRevD.26.1394,
  title = {Covariant calculations at finite temperature: The relativistic plasma},
  author = {Weldon, H. Arthur},
  journal = {Phys. Rev. D},
  volume = {26},
  issue = {6},
  pages = {1394--1407},
  numpages = {0},
  year = {1982},
  month = {Sep},
  publisher = {American Physical Society},
  doi = {10.1103/PhysRevD.26.1394},
  url = {https://link.aps.org/doi/10.1103/PhysRevD.26.1394}
}

@article{PhysRevD.9.3320,
  title = {Symmetry behavior at finite temperature},
  author = {Dolan, L. and Jackiw, R.},
  journal = {Phys. Rev. D},
  volume = {9},
  issue = {12},
  pages = {3320--3341},
  numpages = {0},
  year = {1974},
  month = {Jun},
  publisher = {American Physical Society},
  doi = {10.1103/PhysRevD.9.3320},
  url = {https://link.aps.org/doi/10.1103/PhysRevD.9.3320}
}

@article{PhysRevD.8.4332,
  title = {Improved Treatment for the Infrared-Divergence Problem in Quantum Electrodynamics},
  author = {Grammer, G. and Yennie, D. R.},
  journal = {Phys. Rev. D},
  volume = {8},
  issue = {12},
  pages = {4332--4344},
  numpages = {0},
  year = {1973},
  month = {Dec},
  publisher = {American Physical Society},
  doi = {10.1103/PhysRevD.8.4332},
  url = {https://link.aps.org/doi/10.1103/PhysRevD.8.4332}
}

@article{PhysRev.110.974,
  title = {Bremsstrahlung of Very Low-Energy Quanta in Elementary Particle Collisions},
  author = {Low, F. E.},
  journal = {Phys. Rev.},
  volume = {110},
  issue = {4},
  pages = {974--977},
  numpages = {0},
  year = {1958},
  month = {May},
  publisher = {American Physical Society},
  doi = {10.1103/PhysRev.110.974},
  url = {https://link.aps.org/doi/10.1103/PhysRev.110.974}
}

@article{PhysRevLett.20.86,
  title = {Extension of the Low Soft-Photon Theorem},
  author = {Burnett, T. H. and Kroll, Norman M.},
  journal = {Phys. Rev. Lett.},
  volume = {20},
  issue = {2},
  pages = {86--88},
  numpages = {0},
  year = {1968},
  month = {Jan},
  publisher = {American Physical Society},
  doi = {10.1103/PhysRevLett.20.86},
  url = {https://link.aps.org/doi/10.1103/PhysRevLett.20.86}
}

@article{FRENKEL1984231,
title = {Non-abelian eikonal exponentiation},
journal = {Nuclear Physics B},
volume = {246},
number = {2},
pages = {231-245},
year = {1984},
issn = {0550-3213},
doi = {https://doi.org/10.1016/0550-3213(84)90294-3},
url = {https://www.sciencedirect.com/science/article/pii/0550321384902943},
author = {J. Frenkel and J.C. Taylor},
}

@article{Bauer_2002,
   title={Soft-collinear factorization in effective field theory},
   volume={65},
   ISSN={1089-4918},
   url={http://dx.doi.org/10.1103/PhysRevD.65.054022},
   DOI={10.1103/physrevd.65.054022},
   number={5},
   journal={Physical Review D},
   publisher={American Physical Society (APS)},
   author={Bauer, Christian W. and Pirjol, Dan and Stewart, Iain W.},
   year={2002},
   month=feb }

@article{Beneke:2002ph,
    author = "Beneke, M. and Chapovsky, A. P. and Diehl, M. and Feldmann, T.",
    title = "{Soft collinear effective theory and heavy to light currents beyond leading power}",
    eprint = "hep-ph/0206152",
    archivePrefix = "arXiv",
    reportNumber = "PITHA-02-09",
    doi = "10.1016/S0550-3213(02)00687-9",
    journal = "Nucl. Phys. B",
    volume = "643",
    pages = "431--476",
    year = "2002"
}

@article{BENEKE1998321,
title = {Asymptotic expansion of Feynman integrals near threshold},
journal = {Nuclear Physics B},
volume = {522},
number = {1},
pages = {321-344},
year = {1998},
issn = {0550-3213},
doi = {https://doi.org/10.1016/S0550-3213(98)00138-2},
url = {https://www.sciencedirect.com/science/article/pii/S0550321398001382},
author = {M. Beneke and V.A. Smirnov},

}

@article{PhysRevLett.108.151601,
  title = {Rapidity Renormalization Group},
  author = {Chiu, Jui-yu and Jain, Ambar and Neill, Duff and Rothstein, Ira Z.},
  journal = {Phys. Rev. Lett.},
  volume = {108},
  issue = {15},
  pages = {151601},
  numpages = {5},
  year = {2012},
  month = {Apr},
  publisher = {American Physical Society},
  doi = {10.1103/PhysRevLett.108.151601},
  url = {https://link.aps.org/doi/10.1103/PhysRevLett.108.151601}
}

@article{PhysRevD.76.074002,
  title = {The zero-bin and mode factorization in quantum field theory},
  author = {Manohar, Aneesh V. and Stewart, Iain W.},
  journal = {Phys. Rev. D},
  volume = {76},
  issue = {7},
  pages = {074002},
  numpages = {54},
  year = {2007},
  month = {Oct},
  publisher = {American Physical Society},
  doi = {10.1103/PhysRevD.76.074002},
  url = {https://link.aps.org/doi/10.1103/PhysRevD.76.074002}
}

@article{BECHER201241,
title = {Analytic regularization in Soft-Collinear Effective Theory},
journal = {Physics Letters B},
volume = {713},
number = {1},
pages = {41-46},
year = {2012},
issn = {0370-2693},
doi = {https://doi.org/10.1016/j.physletb.2012.05.016},
url = {https://www.sciencedirect.com/science/article/pii/S0370269312005333},
author = {Thomas Becher and Guido Bell}
}

@article{ECHEVARRIA2013795,
title = {Soft and collinear factorization and transverse momentum dependent parton distribution functions},
journal = {Physics Letters B},
volume = {726},
number = {4},
pages = {795-801},
year = {2013},
issn = {0370-2693},
doi = {https://doi.org/10.1016/j.physletb.2013.09.003},
url = {https://www.sciencedirect.com/science/article/pii/S037026931300720X},
author = {Miguel G. Echevarría and Ahmad Idilbi and Ignazio Scimemi}
}

@article{NIEMI1984105,
title = {Finite-temperature quantum field theory in Minkowski space},
journal = {Annals of Physics},
volume = {152},
number = {1},
pages = {105-129},
year = {1984},
issn = {0003-4916},
doi = {https://doi.org/10.1016/0003-4916(84)90082-4},
url = {https://www.sciencedirect.com/science/article/pii/0003491684900824},
author = {A.J Niemi and G.W Semenoff}
}

@article{Kobes1985,
  author  = {Kobes, R. L. and Semenoff, G. W. and Weiss, N.},
  title   = {Real-time Feynman rules for gauge theories with fermions at finite temperature and density},
  journal = {Zeitschrift f{\"u}r Physik C: Particles and Fields},
  year    = {1985},
  volume  = {29},
  number  = {3},
  pages   = {371--380},
  doi     = {10.1007/BF01565184},
  url     = {https://doi.org/10.1007/BF01565184},
  issn    = {1431-5858}
}

@article{GATHERAL198390,
title = {Exponentiation of eikonal cross sections in nonabelian gauge theories},
journal = {Physics Letters B},
volume = {133},
number = {1},
pages = {90-94},
year = {1983},
issn = {0370-2693},
doi = {https://doi.org/10.1016/0370-2693(83)90112-0},
url = {https://www.sciencedirect.com/science/article/pii/0370269383901120},
author = {J.G.M. Gatheral},
}

@article{PhysRev.140.B516,
  title = {Infrared Photons and Gravitons},
  author = {Weinberg, Steven},
  journal = {Phys. Rev.},
  volume = {140},
  issue = {2B},
  pages = {B516--B524},
  numpages = {0},
  year = {1965},
  month = {Oct},
  publisher = {American Physical Society},
  doi = {10.1103/PhysRev.140.B516},
  url = {https://link.aps.org/doi/10.1103/PhysRev.140.B516}
}

@article{PhysRevLett.102.162001,
  title = {Infrared Singularities of Scattering Amplitudes in Perturbative QCD},
  author = {Becher, Thomas and Neubert, Matthias},
  journal = {Phys. Rev. Lett.},
  volume = {102},
  issue = {16},
  pages = {162001},
  numpages = {4},
  year = {2009},
  month = {Apr},
  publisher = {American Physical Society},
  doi = {10.1103/PhysRevLett.102.162001},
  url = {https://link.aps.org/doi/10.1103/PhysRevLett.102.162001}
}

\end{document}